# Über wissenschaftliche Exzellenz und Wettbewerb

Gerald Schweiger, Technische Universität Graz, Österreich

Das Streben nach Exzellenz scheint das True North der akademischen Welt zu sein – auch in Österreich. Die Universität Innsbruck verweist in ihrem Entwicklungsplan auf zwei zentrale strategische Ziele: *„Exzellenz in der Lehre sowie Exzellenz in der Forschung"* [1], im Mission Statement der TU Wien liest man *„[...] mit einem klaren Bekenntnis zur Exzellenz"* [2] und in der Internationalisierungsstrategie der FH Burgenland *„[...] um Exzellenz in Lehre, Forschung und Verwaltung sicherzustellen"* [3]. Das berufsbegleitende MBA Programm der WU Wien zeichnet sich durch *„eine Kombination aus herausragender akademischer Exzellenz, [...]"* [4] mit anderen Dingen aus. Was ist mit Exzellenz gemeint? Kann Exzellenz gemessen werden? Oder steht Exzellenz einfach auf der Checkliste von Wörtern für solche Texte: Diversität, Inklusion, Exzellenz,...

Auch Förderprogramme setzen auf Exzellenz. Der Fonds zur Förderung der wissenschaftlichen Forschung (FWF) hat die Exzellenzinitiative excellent=austria ins Leben gerufen [5]; die österreichische Forschungsförderungsgesellschaft (FFG) hat eine eigene Förderschiene mit dem Namen Competence Centers for Excellent Technologies [6]. Aber auch für den engeren Kontext der Forschung gilt: Was ist mit exzellenter Forschung gemeint und wie wird sie gemessen?

In der akademischen Literatur wird Exzellenz als ein unscharfer und mehrdeutiger Begriff beschrieben [7], [8]. Die Unschärfe des Exzellenzbegriffs wird jedoch unterschiedlich interpretiert. Einige sehen in darin eine Stärke, die eine Zusammenarbeit zwischen Akteuren aus verschiedenen Bereichen mit unterschiedlichen Auffassungen von Exzellenz ermöglicht [9], [10]. Andere sehen darin eine Schwäche und Exzellenz wird als rationalisierender Mythos beschrieben, *„performing a rhetorical function but lacking any intrinsic meaning"* [11]. Trotz dieser Unschärfe hat die wissenschaftliche Gemeinschaft verschiedene Metriken vorgeschlagen, um Exzellenz in der Forschung zu quantifizieren; in den meisten Fällen handelt es sich um bibliometrische Indikatoren wie die Anzahl an Publikationen oder Zitaten.

## 1 Exzellenz und Wettbewerb

Die Wissenschaft ist ein komplexes System, das mit anderen komplexen Systemen in Wechselwirkung steht [12]; nach Bruno Latour ist sie ein System, in dem Wissenschaftler:innen und deren Institutionen *„knit, weave and knot together into an overarching scientific fabric"* [13]. Die Verteilung von Forschungsgeldern ist ein wichtiger Baustein dieses komplexen Systems. Vereinfacht kann zwischen zwei Arten der Verteilung unterschieden werden: Entweder konkurrieren Wissenschaftler:innen aktiv um Forschungsgelder (z. B. durch das Schreiben von Forschungsanträgen), oder sie erhalten Gelder, ohne aktives Zutun (z. B. durch direkte Finanzierung der Universitäten).

Nehmen wir um der Argumentation willen an, dass Exzellenz anhand bibliometrischer Indikatoren gemessen werden kann. Erhöht Wettbewerb die Exzellenz oder ist er eher hinderlich? Mit den von Latour beschriebenen Eigenschaften der Wissenschaft – technisch würde man von der Nichtlinearität, Multidimensionalität und Selbstorganisation komplexer Systeme sprechen – ist die Wissenschaft ein Beispiel für Systeme, in denen es äußerst schwierig ist, aus nicht-experimentellen Daten kausale Schlüsse zu ziehen[1].

In einer Studie wurde in 17 Ländern der Einfluss kompetitiver Forschungsförderung auf die Effizienz analysiert, wobei Effizienz als Änderung der Fördergelder im Vergleich zur Änderung hochzitierter Veröffentlichungen definiert wurde [14]. Die Daten zeigen eine negative Korrelation von 0,3 zwischen der Effizienz und dem Grad kompetitiver Förderung. Wettbewerb wirkt sich laut dieser Analyse negativ auf die Effizienz aus. Die Autoren weisen darauf hin, dass die Ergebnisse aufgrund der geringen Datenmenge mit Vorsicht interpretiert werden sollten; ich bin der Meinung, dass die Korrelation einfach zufällig sein kann. In einer anderen Studie wurden Daten aus acht Ländern ausgewertet, um die Auswirkung kompetitiver Forschungsförderung auf die Produktivität - gemessen an der Anzahl an Publikationen - zu analysieren [15]. Länder mit hohem Wettbewerb (z.B. Großbritannien) sind effizient, konnten die Effizienz jedoch nicht verbessern; andere Länder mit weniger Wettbewerb sind entweder fast genauso effizient (z.B. Dänemark) oder konnten ihre Effizienz trotz relativ geringem Wettbewerb erhöhen (z.B. Schweden).

## 2  Gewinner und Verlierer

Ausgehend von einem ORF-Beitrag über Wettbewerb und Innovation gab der FWF eine Stellungnahme ab, in der auf eine Erhebung verwiesen wird, die zeigt,

> *„dass Forscherinnen und Forscher aus Österreich ohne Förderungen der großen Förderorganisationen, wie FWF oder European Research Council, weniger zitiert wurden als jene, die Förderungen und Grants erhielten. 17 Zitationen pro Publikation von nicht geförderten Forschenden stehen 33 Zitationen bei geförderten Projekten gegenüber"* [16].

Vier kritische Überlegungen zu dieser Aussage.
<u>Erstens:</u> In einer Studie wurde ein Vergleich zwischen Forscher:innen mit bewilligten und jenen mit abgelehnten Projekten in der Biologie und den Sozialwissenschaften anhand bibliometrischer Indikatoren durchgeführt [17]. In beiden Disziplinen erzielten diejenigen mit bewilligten Projekten durchschnittlich bessere Ergebnisse als ihre Kolleg:innen mit abgelehnten Projekten. Werden jedoch nur die besten abgelehnten Forscher:innen berücksichtigt (die gleiche Anzahl wie geförderte Forscher:innen), schneiden diese besser ab als jene mit bewilligten Projekten[2].

---

[1] Der Statistiker Freedman sagt dazu *"[…] Causal inferences can be drawn from nonexperimental data. However, no mechanical rules can be laid down for the activity. […] Instead, causal inference seems to require an enormous investment of skill, intelligence, and hard work. Many convergent lines of evidence must be developed. Natural variation needs to be identified and exploited. Data must be collected. Confounders need to be considered. Alternative explanations have to be exhaustively tested"* [49].

Vielleicht kann ein Auswahlverfahren auf der Grundlage von Peer-Reviews und Panel-Entscheidungen sehr schlechte Projekte aussortieren, aber es scheint kein geeignetes Instrument zu sein, um eine nachvollziehbare Auswahl unter den verbleibenden Projekten zu treffen.

> „The best rejected proposals score on average as high on the outcomes of the peer-review process as the accepted proposals" [18]; "Some of the losing proposals are truly bad, but not all; many of the rejected proposals are no worse than many of the funded ones . […] When proposals are abundant and money is scarce, the vast majority of putative funding errors are exclusory; a large number of proposals are rejected that are statistically indistinguishable from an equal number accepted" [19].

<u>Zweitens:</u> Auch wenn es Disziplinen gibt, in denen Forscher:innen mit geförderten Projekten bessere bibliometrische Indikatoren aufweisen, kann dies auf verschiedene Gründe zurückzuführen sein - einige davon sind trivial [20]: (i) Bessere Forscher:innen sind erfolgreicher bei der Einwerbung von Fördermitteln; (ii) durch Fördermittel können Forscher:innen bessere Versionen ihrer geplanten Forschung durchführen, z.B. durch größere empirische Umfragen oder neueres Equipment. (iii) Die Fördermittel könnten Studien ermöglichen, die ohne externe Finanzierung unmöglich wären.
Selbst wenn die Daten zeigen würden, dass Forscher:innen mit geförderten Projekten bessere bibliometrische Indikatoren aufweisen, kann nicht einfach von einer Ursache-Wirkung-Beziehung ausgegangen werden

<u>Drittens:</u> Auch wenn es Disziplinen gibt, in denen Forscher:innen mit geförderten Projekten bessere bibliometrische Indikatoren aufweisen, kann daraus nicht geschlossen werden, dass wettbewerbsorientierte Systeme insgesamt besseren Output (gemessen an bibliometrische Indikatoren) hervorbringen als beispielsweise aleatorische oder egalitäre Systeme. Es muss gezeigt werden, dass die Kosten des Wettbewerbs gerechtfertigt sind (siehe dazu das Kapitel 4).

<u>Viertens:</u> In der Stellungnahme des FWF wird auf eine Erhebung verwiesen, in der quer über alle Disziplinen die durchschnittliche Anzahl von Zitaten für Artikel im Zeitraum zwischen 2013 und 2022 analysiert werden. Zwei Probleme: (i) Zitate sind zeit- und fachabhängig [21] und (ii) es können nicht einfach Artikel, die im Rahmen von Forschungsprojekten entstanden sind, mit Artikeln verglichen werden, die ohne Förderung entstanden sind [22].

---

[2] *Es wurden zwei Vergleiche angestellt: (i) 119 geförderte Projekte im Vergleich zu allen 251 abgelehnten Projekten; (ii) 119 geförderte Projekte im Vergleich zu den besten 119 abgelehnten Projekten.*

Neben einer Input (Geld rein) und Output (Publikationen raus) Analyse gibt es noch weitere Faktoren, die bei der Bewertung von Wettbewerb in der Verteilung von Fördermitteln berücksichtigen werden sollten.

## 3 Sind Förderentscheidungen zuverlässig?

Die Zuverlässigkeit von Auswahlprozessen wird hinsichtlich der Variabilität der Peer-Review und Panel-Bewertungen analysiert. Einige Studien zeigen keine Übereinstimmung bei der Bewertung derselben Anträge durch verschiedene Gutachter:innen [23]; andere Studien zeigen eine sehr geringe [24] bis moderate Übereinstimmung [25]. Im Bereich der Panel Entscheidungen zeigten Cole und Kollegen 1981 erstmals, dass *„getting a research grant depends to a significant extent on chance"* [26]. Eine aktuellere Studie analysierte 2705 medizinische Forschungsanträge, die in Australien eingereicht wurden [27]. Die Autoren kommen zu dem Schluss, dass der Auswahlprozess kostspielig und nicht zuverlässig ist. Dass der Zufall im Begutachtungsprozess eine Rolle spielt, wurde auch in qualitativen Studien bestätigt [28].

Ich möchte noch eine Hypothese aufstellen: Die Zuverlässigkeit von Entscheidungsprozessen ist in der angewandten Forschung noch geringer als in der Grundlagenforschung. In den Grundlagenwissenschaften kann sehr genau definiert werden, welche Personen für Gutachten und Panels qualifiziert sind. In der angewandten Forschung muss beispielsweise ein Maschinenbauingenieur beurteilen, ob probabilistische neuronale Netze ein geeignetes Werkzeug für die Diagnose von Windkraftanlagen sind; neben der wissenschaftlichen Exzellenz müssen dann noch weitere Kriterien wie das Marktpotenzial oder der Beitrag zur Technologieführerschaft eines Landes beurteilt werden. Auf Seiten der Antragsteller:innen führt dies zu geschönten, mit Buzzwords überladenen Forschungsanträgen (insbesondere in Systemen ohne starke ex-post Evaluierung) und auf Seiten der Begutachtung zu einem Auswahlverfahren, dessen Zuverlässigkeit sich kaum von einer klassischen Münzwurfentscheidung unterscheiden lässt.

## 4 Die Kosten des Wettbewerbs

In wettbewerbsorientierten Systemen müssen die ausgeschütteten Fördergelder die Kosten für die Verteilung deutlich übersteigen. Zu den Kosten für die Verteilung zählen die Gehälter der Forscher:innen, die für das Verfassen der Anträge verantwortlich sind, Kosten für Gutachten und Panel-Entscheidungen sowie Verwaltungskosten. Eine Studie in Australien zeigt, dass sich die Kosten der Verteilung in wettbewerbsorientierten Systemen wie folgt aufteilen: 85 % der Kosten entfallen auf die Antragsteller:innen, 10 % auf Gutachten und Panels und die restlichen 5 % auf die Verwaltung[3] [27]. Aus Sicht des Geldgebers ist die Nettorendite als Funktion der Förderquote eine wichtige Größe [29].

---

[3] Die Verwaltungskosten der größten deutschen Förderorganisation DFG liegen bei 2,3% [29]. Für Österreich gibt es keine offiziellen Zahlen.

Eine Nettorendite von Null entspricht einer Situation, in der alle Mittel ausschließlich für die Einwerbung und Verteilung von Fördermitteln verwendet werden - es wird keine einzige Stunde tatsächlicher Forschung finanziert. Die Nettorendite hängt hauptsächlich von er Förderquote ab - der Punkt der Nettorendite von Null wird bei Förderquoten überschritten, die in der derzeitigen Forschungslandschaft nicht unüblich sind [29].

Warum sind die Kosten der Antragsstellung so hoch? Das Schreiben von Forschungsanträgen erfordert viel Zeit. Studien in verschiedenen Disziplinen zeigen, dass das Schreiben eines einzelnen Antrags etwa 25-50 Tage in Anspruch nimmt [30]–[32]. Bei durchschnittlichen Akzeptanzraten zwischen 10-25% ergibt das 100 bis 500 Personentage für ein gefördertes Projekt. Eine Studie der Förderung medizinischer Forschung in Australien zeigt, dass Forscher:innen im Jahr 2013 rund 550 Arbeitsjahre investierten, um Forschungsanträge zu schreiben; dies entspricht 41 Millionen Euro an Gehältern und 14% des gesamten Budgets für medizinische Forschung [31]. Eine Evaluierung der europäischen H2020-Programme ergab, dass zwischen 30 und 50 % der Mittel, die die Länder aus Horizont 2020 erhalten, für die Erstellung der Anträge ausgegeben werden [33]. Eine Evaluierung des britischen Research Councils zeigt, dass die Gutachter:innen in der Periode 2005/06 rund 192 Jahre mit der Begutachtung von Anträgen verbracht haben [34].

Eine Studie über staatlich geförderte Forschungsprojekte in den USA zeigt, dass Principal Investigators im Durchschnitt rund 45% ihrer Zeit für administrative Tätigkeiten im Zusammenhang mit der Beantragung und Verwaltung von Projekten aufwenden, anstatt aktive Forschung zu betreiben [35]. Eine Studie in Österreich zeigt, dass 90% der Forscher:innen der Meinung sind, dass sie zu viel Zeit für das Schreiben von Forschungsanträgen aufbringen; nur 10% der Forscher:innen glauben, dass sich wettbewerbsorientierte Fördersysteme positiv auf die Qualität der Forschung auswirken [30].

Oftmals sind erfolglose Anträge jedoch keine komplette Zeitverschwendung; sie haben auch positive externe Effekte: *„Scientists can generate, refine, and share their ideas, regardless of whether or not they ultimately receive the funding"* [36]. So hat eine Studie gezeigt, dass 39 % der Zeit, die Forscher für die Vorbereitung von Anträgen aufwenden, ihrer eigentlichen Forschung zugute kommt [35].

## 5 Diskussion

<u>Erstens:</u> Es braucht einen offenen Wettbewerb der Ideen, wie Fördergeld verteilt werden soll; zwei in den letzten Jahren mehrfach diskutierte Optionen sind die aleatorische (fund at random) und egalitäre Verteilung (fund everybody) [37]. Aktuelle Studien zeigen, dass eine Grundfinanzierung bei gleichmäßiger Verteilung von der akademischen Gemeinschaft positiv gesehen wird [38], selbst unter „high-performing" Wissenschafter:innen [39].

Um verschiedene Systeme zu vergleichen, braucht es (randomisierte) Experimente, quantitative Modelle und offen zugängliche Daten von Förderorganisationen. Diese Vergleiche müssen für verschiedene Disziplinen und entlang des Kontinuums zwischen Grundlagenforschung und angewandter Forschung angestellt werden. Für quantitative Modelle werden verschiedene Daten benötigt: (i) Daten über Entscheidungsprozesse. Stimmen die Bewertungen der Gutachten überein? Wie hoch ist die Übereinstimmung an geförderten Projekten, wenn unterschiedliche Panels basierend auf den gleichen oder unterschiedlichen Gutachten Entscheidungen treffen?[4] (ii) Bibliometrische Daten der abgelehnten und akzeptierten Antragsteller:innen. (iii) Administrative Kosten für die verschiedenen Verteilsysteme. (iv) Daten über die Zeit, die für das Antragschreiben benötigt wird und eine Abschätzung wie viel davon der eigentlichen Forschung der Antragsteller:innen zugutekommt. Modelle könnten beispielsweise auch um die Dimension der sozialen Kosten erweitert werden [40][5]. So kann der Einfluss des Wettbewerbs auf toxische Forschungskulturen [35] oder auf das (Quiet) Quitting in der Wissenschaft [36] berücksichtigt werden und wie sich das wiederum im Zeitverlauf auf den wissenschaftlichen Output auswirkt.

Zweitens: Es braucht einen offenen und transparenten Diskurs über Exzellenz. Der Archetyp des modernen Spitzenforschers: Er orchestriert eine Maschinerie, die den Output an Paper (oder Zitaten)[6] und den Input an Drittmitteln maximiert [41]. Die Erweiterung des Wissens bleibt dann (wenn überhaupt) nur als sekundäres Ziel. Die eigentliche Forschung wird weitgehend von Assistent:innen und Doktorand:innen durchgeführt, denn für diejenigen, die an der Spitze der Hierarchie stehen, sind die Opportunitätskosten für die Forschung meist zu hoch. Es ist sowohl für ihre eigene Karriere als auch für die Institution von Vorteil, wenn sie die oben genannten Maschinerie am Laufen halten [42]. Die Annahme, dass das Wissen einfach proportional zur Anzahl wissenschaftlicher Publikationen steigt, ist zweifelhaft [43]. So heißt es beispielsweise in einem Artikel über die Forschung im Bereich der Nachhaltigkeit: *„up to 50% of the articles that are now being published in many interdisciplinary sustainability and transitions journals may be categorized as scholarly bullshit"* [44]. Ein Grund für diese Situation ist der Wettlauf um Paper, Zitate und Forschungsgelder; oder wie Binswanger es formuliert: *„Taste for Science" [...] which is based on intrinsic motivation and supposed to guide scientists was replaced by the extrinsically motivated ''Taste for Publications"* [42].

Der Präsident des FWF Christof Gattringer hat in einer Radiodiskussion den interessanten Gedanken über Exzellenz formuliert: *„Exzellenz ist das, was in 20 Jahren im Lehrbuch steht"* [45]. Ich würde ergänzen: *„was das Potenzial hat"*, in 20 Jahren im Lehrbuch zu stehen. Dazu Zeilinger:

> *„Die Wissenschaft ist nicht planbar, doch das geht immer mehr verloren. Im Forschungsbetrieb wird immer mehr Planung verlangt und Kennzahlen, an denen man beurteilen kann, ob man die Pläne erreicht hat."* [46]; *„es geht darum das Ungewöhnliche zu finden und für das*

---

[4] Wichtig ist anzumerken: diese Daten beziehen sich auf die Konsistenz der Entscheidung – nicht auf deren Richtigkeit (gemessen an bibliometrischen Indikatoren oder anderen Metriken).
[5] Siehe dazu auch den Beitrag von Stephan Pühringer „Wie viel Wettbewerb wollen wir (uns leisten)?"
[6] Das Problem der unethischen Autorenschaft [50] ist eines von vielen damit verbundenen Problemen.

*Unvorhersehbare offen zu sein […] es geht nicht um den nächsten Schritt den man klar definieren kann. Der nächste offenkundige Schritt ist zu wenig"* [47].

Neben der Förderung von Neuem muss ein System aber meiner Meinung nach noch etwas sicherstellen: Es muss Rahmenbedingungen schaffen, in denen schlechte Ideen so schnell wie möglich verworfen werden können und so wenig wie möglich in wissenschaftliche Sackgassen investiert wird. Dieser Punkt erscheint in der aktuellen Diskussion zu wenig berücksichtig zu werden. Wie schnell können wissenschaftliche Sackgassen verlassen werden, wenn Forscher:innen oder ganze Organisationen von kompetitiven Drittmitteln abhängig sind, die in einem Zwei- bis Dreijahreszyklus erneuert werden müssen?

<u>Drittens:</u> Wir sollten – vor allem in Österreich – über den Stellenwert der Grundlagenforschung diskutieren; und über das Verhältnis von Fördermittel für angewandte und Grundlagenforschung; und darüber, ob Zeilinger Recht hat, wenn er meint, dass *„einiges, was unter angewandter Forschung geführt wird, vielleicht nicht ganz Forschung ist"* [48]. Bei vielen angewandten Projektausschreibungen muss rezeptähnlich angegeben werden, wie, wann und mit welchen Methoden Ziele erreicht werden. Ist das noch Forschung?